\documentclass[12pt]{article}
\usepackage{epsfig,amsmath,amsfonts}%

\def \beq {\begin{equation}}
\def \eeq {\end{equation}}
\def \beqar {\begin{eqnarray}}
\def \eeqar {\end{eqnarray}}
\def \beqa* {\begin{eqnarray*}}
\def \eeqa* {\end{eqnarray*}}
\def \pa {\partial}
\def \<< {\langle}
\def \>> {\rangle}

\def \d {\,{\rm{d}}}
\def \polytrop {\nu}

\begin{document}
\title{Dark Energy and Dark Matter unification \\via superfluid Chaplygin gas}%
\author{V.\,A.~Popov}%
\date{}

\maketitle%

\begin{center}\small \textit{Department of General Relativity and
Gravitation,\\Kazan State University,\\Kremlyovskaya st. 18, Kazan
420008, Russia}\end{center}

\begin{center}\small Email address: vladipopov@mail.ru\\
Tel: +7(843)520 77 42
\end{center}

\begin{abstract}
A new model describing the dark sector of the universe is
established. The model involves Bose-Einstein condensate (BEC) as
dark energy (DE) and an excited state above it as dark matter
(DM). The condensate is assumed to have a negative pressure and is
embodied as an exotic fluid with Chaplygin equation of state.
Excitations are described as a quasiparticle gas. It is shown that
the model is not in disagreement with the current observations of
the cosmic acceleration. The model predicts increase of the
effective cosmological constant and a complete disappearance of
the matter at the far future.
\end{abstract}

\vspace*{12pt}

\noindent{\small \emph{Keywords:} accelerated expansion, Dark
Energy, Dark Matter, relativistic superfluid, Chaplygin gas

\noindent\emph{PACS:} 95.36.+x,  95.35.+d, 04.40.-b, 98.80.Jk,
47.37.+q}

%

\newpage

\section{Introduction}

The energy content of the Universe is a fundamental issue in
cosmology. Observational data are evidence of accelerating flat
Friedmann-Robertson-Walker model, constituted of about 1/4 of
baryonic and dark matter  and about 3/4 of a dark energy
component. The DM content was originally inferred from spiral
galactic rotation curves and then was supported by gravitational
lensing  and cosmic microwave background observations.

The essential feature of DE is that its pressure must be negative
to reproduce the present accelerated cosmic expansion. The
simplest DE model, the cosmological constant, is indeed the vacuum
energy with the equation of state $p=-\rho$. The models for which
$p<-\rho$ has been denoted phantom energy, and possesses peculiar
properties, such as, an infinitely increasing energy density
\cite{Caldwell1}, negative temperatures \cite{Nojiri1}, and the
violation of the null energy condition. A number of models, such
as quintessence \cite{Wetterich1} and k-essence
\cite{Armendariz1}, are based on scalar field theories. These
models are parameterized by an equation of state $p<-\rho/3$. For
a recent review of DE models and references see \cite{Copeland1}.

An alternative model is that of the Chaplygin gas, also denoted as
quartessence, based on a negative pressure fluid, which is
inversely proportional to the energy density \cite{Kamenshchik1}.
The equation of state representing the generalized Chaplygin gas
(GCG) is given by $p_{\rm Ch}= - A/\rho^\alpha_{\rm Ch}$ with
positive constants $A$ and $\alpha$ ($0<\alpha\le 1$)
\cite{Bento1}. An attractive feature of these models, is that at
early times, the energy density behaves as a matter, $\rho_{\rm
Ch}\propto a^{-3}$, where $a$ is the scale factor, and as a
cosmological constant at a later stage, $\rho_{\rm Ch}=$const. It
is also suggested that at an intermediate stage the energy density
$\rho_{\rm Ch}$ consists of both vacuum and soft matter (matter
with the equation of state $p=\alpha\rho$) contributions. This is
favorable to use the GCG model for a DE and DM unification
\cite{Bento1,Bilic1,Bento2}.

Some different approach to the same problem is realized in this
work. The feature of the scalar field used in the  model
\cite{Bilic1} is spontaneous symmetry breaking that leads to a
nonzero expectation value of the field. In other words,
Bose-Einstein condensation of the scalar field into the state with
zero momentum takes place. Similar effect holds in the theories of
superconductivity and superfluidity. At zero temperature
superfluid is in its ground state. If $T\ne 0$ particle-like
excitations arises above the ground state. In this case the system
can be divided into a background superfluid condensate and a
quasiparticle gas. This separation naturally leads to two-fluid
dynamics in which the condensate is said to be a superfluid
component and the quasiparticle gas forms a normal component.

The model proposed  in this letter represents the dark sector of
the universe as a superfluid where the superfluid condensate is
considered as DE and the normal component is interpreted as DM. To
provide the accelerated expansion the potential of the scalar
field must have specific form and entail a negative pressure of
the superfluid background.

We base our analysis on the action
\beq\label{Action}
S=\int \left(-\frac{R}{16\pi G}+{\cal L}\right)\sqrt{-g}\d^4x,
\eeq
where
Lagrangian ${\cal L}$ %
associated with a generalized hydrodynamic pressure function
depends only on one variable if we consider pure condensate, and
on three variables when we include the excitation gas.

\section{Condensate in WKB-approximation}

Consider a system with spontaneously breaking of $U(1)$ symmetry
described by a complex scalar field $\hat\psi$ with nonzero Gibbs
expectation $\phi(x)=\langle\hat\psi\rangle$, the condensate wave
function. Quantum fluctuations $\hat\psi'=\hat\psi-\phi$ can be
considered in terms of quasiparticles.

If the quasiparticle gas is dilute or interaction between the
quasiparticles and the condensate is weak then influence of the
elementary excitations on the ground state can be neglected. In
this case the condensate is described by the Lagrangian
\beq\label{LagrangianBasic}
{\cal L} = \pa_\nu \phi^* \pa^\nu \phi - V(\phi^*\phi).
\eeq

It is useful for further study to represent the Lagrangian
(\ref{LagrangianBasic}) in an equivalent hydrodynamic form. For
this purpose we write the condensate wave function in terms of
modulus and phase:
\beq\label{ExpForm}
\phi=\frac{\sigma}{\sqrt{2}}e^{-i\chi}.
\eeq
Substituting (\ref{ExpForm}) into the field equation, the real and
imaginary parts yield respectively
\beq\label{EilerEquReal}
\nabla_\nu\nabla^\nu\sigma+\sigma\left(2\frac{\d
V}{\d\sigma^2}-\pa_\nu\chi\pa^\nu\chi\right)=0,
\eeq
and
\beq\label{CurrentConservation}
\nabla_\nu(\sigma^2 \pa^\nu \chi)=0.
\eeq

Equation (\ref{CurrentConservation}) is a conservation law for the
4-current
\beq
j_\nu= i\left( \phi^*\pa_\nu\phi - \pa_\nu\phi^*\phi
\right)=\sigma^2 \pa_\nu \chi.
\eeq
The gradient of the condensate phase is a superfluid momentum
which can be written in terms of a unit 4-vector $V_\nu$:
\beq
\pa_\nu \chi = \mu_\nu =\mu V_\nu,
\eeq
where $\mu$ is a chemical potential. In the present context
$j_\nu$ is the particle current, therefore
\beq
\sigma^2=\frac{n_{\rm c}}{\mu},
\eeq
where $n_{\rm c}$ is a particle density of BEC.

If the modulus $\sigma$ varies slower than the phase $\chi$,
$\pa_\nu\sigma\le \sigma\pa_\nu\chi$, then we can neglect
derivatives $\pa_\nu\sigma$ that corresponds to the WKB expansion
of the condensate wave function up to the first
order%
.

In this approximation equation (\ref{EilerEquReal}) takes the form
\beq
\sigma\mu^2-\frac{\d V}{\d\sigma}=0,
\eeq
and the particle density is
\beq\label{NumberDensityGeneral}
n_{\rm c}^2=\sigma^3\frac{\d V}{\d\sigma}.
\eeq

The energy-momentum tensor is found as variation of the Lagrangian
with respect to the metric:
\beq
T_{\mu\nu} = \frac{2}{\sqrt{-g}} \frac{\delta \left(\sqrt{-g}
{\cal L}\right)}{\delta g^{\mu\nu}} = \sigma^2 \pa_\mu\chi
\pa_\nu\chi  - g_{\mu\nu} \left( \frac{\sigma^2}{2}
\pa_\lambda\chi \pa^\lambda\chi + V(\sigma^2)\right) = (\rho_{\rm
c}+p_{\rm c}) V_\mu V_\nu -p_{\rm c}g_{\mu\nu},
\eeq
where the pressure and the energy density
\beq\label{PressureGeneral}
p_{\rm c}=\frac{1}{2}\frac{\d V}{\d\sigma}\sigma - V, \qquad %
\rho_{\rm c}=\frac{1}{2}\frac{\d V}{\d\sigma}\sigma + V
\eeq
depend on only one variable $\mu$.

For various contexts we can apply a rather different kind of
potential. In our case the condensate is assumed to possess a
negative pressure. For this reason let us take the potential
\beq
V(\phi^*\phi)=M \left( \frac{\phi^*\phi}{\lambda} +
\frac{\lambda}{\phi^*\phi} \right).
\eeq
In accordance with equations (\ref{NumberDensityGeneral}) and
(\ref{PressureGeneral}) one finds that
\beq
n_{\rm c}=\frac{2\lambda\mu}{\sqrt{1-\lambda\mu^2/M}},\qquad%
\rho_{\rm c}=\frac{2M}{\sqrt{1-\lambda\mu^2/M}},
\eeq
and the function of the generalized pressure has the form
\beq\label{PressureInCondensate}
P(\mu)=p_{\rm c}=-2M\sqrt{1-\lambda\mu^2/M}
\eeq
For further purpose it is more rational to eliminate the chemical
potential $\mu$ and to express the hydrodynamic quantities via the
energy density:
\beq\label{ChaplyginEOS}
n_{\rm c}=\sqrt{\frac{\lambda}{M}}\sqrt{\rho_{\rm c}^2-4M^2},\qquad%
p_{\rm c}=-\frac{4M^2}{\rho_{\rm c}},\qquad%
\eeq
and the adiabatic speed of sound is
\beq\label{SoundSpeed}
c_{\rm s}^2=\frac{\d p_{\rm c}}{\d\rho_{\rm
c}}=\frac{4M^2}{\rho_{\rm c}^2}.
\eeq
The equation of state (\ref{ChaplyginEOS}) is uniquely proper to
Chaplygin gas suggested recently by Kamenshchik et
al.~\cite{Kamenshchik1} as an alternative to quintessence and
developed by a number of authors for description of the dark
sector of the universe \cite{Bento1,Bilic1}.

In contrast to these works where pressure of Chaplygin gas is
formed by both DE and DM, this model implies that the equation of
state (\ref{ChaplyginEOS}) concerns with only BEC which is
interpreted as DE.

\section{Relativistic superfluid dynamics}

An efficient approach to description of the excited state is
two-fluid hydrodynamics. This theory does not depend on details of
microscopic structure of the quantum liquid and exploits effective
macroscopic quantities. In the theory there exist two independent
flows, the coherent motion of the ground state named a superfluid
component, and a normal component produced  by the quasiparticle
gas. For this reason  it is necessary to increase the number of
independent variables in the generalized pressure
(\ref{PressureInCondensate}) from one to three
\cite{Khalatnikov1,Carter1}. They correspond to three scalar
invariants which can be constructed from the pair of independent
vectors, namely superfluid $\mu_\alpha$ and thermal
$\theta_\alpha$ momentum covectors so that the general variation
of the generalized pressure in a fixed background is
\beq\label{dP1}
\delta P = n^\alpha \delta \mu_\alpha + s^\alpha \delta
\theta_\alpha.
\eeq
The coefficients $n^\alpha$ and $s^\alpha$ are to be interpreted
as particle number and entropy currents correspondingly. By virtue
of its invariance the pressure is given as a function of three
independent variables,  $I_1=\frac{1}{2}\mu_\alpha
\mu^\alpha,\,I_2=\mu_\alpha
\theta^\alpha,\,I_3=\frac{1}{2}\theta_\alpha \theta^\alpha$.
Taking the derivatives of the pressure, one finds
\beq\label{lin}
n^\alpha = \frac{\pa P}{\pa I_1} \mu^\alpha + \frac{\pa P}{\pa I_2} \theta^\alpha, \qquad%
s^\alpha = \frac{\pa P}{\pa I_2} \mu^\alpha + \frac{\pa P}{\pa
I_3} \theta^\alpha.
\eeq

As soon as the generalized pressure is the Lagrangian density in
the action (\ref{Action}) its variation with respect to the metric
gives the energy-momentum tensor
\beq
T_{\alpha\beta}=\frac{\pa P}{\pa I_1}\mu_\alpha \mu_\beta +
\frac{\pa P}{\pa I_2}(\mu_\alpha \theta_\beta + \theta_\alpha
\mu_\beta) +\frac{\pa P}{\pa I_3}\theta_\alpha \theta_\beta -
Pg_{\alpha\beta}.
\eeq

Instead of the thermal momentum $\theta_\alpha$ let us introduce
an inverse temperature vector
$\beta^\alpha=s^\alpha/(s^\beta\theta_\beta)$ which we use as the
independent vector together with the superfluid momentum
$\mu_\alpha$ since they are comoving to the excitation gas and the
condensate respectively. Corresponding unit 4-velocities are
\beq\label{Velocities}
U^\alpha=\frac{\beta^\alpha}{\sqrt{\beta^\beta \beta_\beta}},
\qquad V^\alpha=\frac{\mu^\alpha}{\sqrt{\mu^\beta \mu_\beta}}.
\eeq

In place of the scalars $I_1,\ I_2,\ I_3$ we use new three
invariants, a chemical potential $\mu=\sqrt{\mu^\beta \mu_\beta}$,
scalar $\gamma=V_\alpha U^\alpha$ associated with the relative
motion of the components, and inverse temperature with respect to
the reference frame comoving to the excitation gas
$\beta=\sqrt{\beta^\beta \beta_\beta}$.

Using (\ref{lin}) and (\ref{Velocities}) the energy-momentum
tensor and the particle number current are readily represented as
\beqar
n^\alpha &=& n_{\rm c}V^\alpha + n_{\rm n}U^\alpha \label{sfPN}\\
T_{\alpha\beta} &=& \mu n_{\rm c} V_\alpha V_\beta + W_{\rm n}
U_\alpha U_\beta - P g_{\alpha\beta},\label{sfEMT}
\eeqar

Relations between the macroscopic quantities involved in equations
(\ref{sfPN}) and (\ref{sfEMT}) are of a quite general form. More
detail information about them can be obtained from statistical
description of the elementary excitations. The quasiparticle
energy spectrum has a significant nonlinear dispersion at high
energy, and therefore completely relativistic description has been
carried out only for a low energy excitations, phonons
\cite{Carter2,Popov1}. Based on the relativistic kinetic theory of
the phonon gas \cite{Popov1} we in particular can obtain
\beq\label{EOSnn}
\frac{\mu n_{\rm n} }{\gamma}=(1-c_{\rm s}^2)W_{\rm n},
\eeq
when phonons prevail over another sorts of quasiparticles.

Let us assume that the generalized pressure function is separated
as follows:
\beq\label{AnsatzSeparatedCondensate}
P(\mu,\beta,\gamma)=p_{\textrm{c}}(\mu)+p_{\textrm{n}}(\mu,\beta,\gamma).
\eeq
Equation (\ref{AnsatzSeparatedCondensate})  is best suited for the
phonon gas since it describes neglect of excitation influence on
the ground state. This ansatz retains equations
(\ref{ChaplyginEOS}) and (\ref{SoundSpeed}) valid for the
condensate in the framework of two-fluid dynamics.

Let us also suppose that the pressure of the normal component
depends on temperature by the power law, $p_{\rm n}\propto
T^{\polytrop+1}$. This assumption is a generalization of the
dependence $p_{\rm n}\propto T^4$ obtained by Carter and Langlois
\cite{Carter2} for the equilibrium distribution of the phonon gas
followed by the relation $p_{\rm n}=c_{\rm s}^2W_{\rm n}/4$. In
our case it transforms to
\beq\label{AnsatzPn}
p_{\rm n}=\frac{c_{\rm s}^2}{1+\polytrop}W_{\rm n}.
\eeq
Equation (\ref{AnsatzPn}) is a kind of barotropic equation of
state and $\polytrop$ is a properly polytropic index. Its value
governs so-called second sound speed $c_2$, the sound speed in the
excitation gas. In the limit of low temperatures when the
quasiparticle contribution becomes small $c_2\to c_{\rm
s}/\sqrt{\polytrop}$, and in the case of the equilibrium phonon
gas it coincides with the result obtained in \cite{Carter2}.

We restrict our consideration to the equation of state
(\ref{AnsatzPn}) for the normal component situated between the
dust one and the stiff one. It is evident from equation
(\ref{AnsatzPn}) that this constraint implies $\polytrop\ge 1$.

\section{Universe with BEC}

\subsection{Equations of motion}

The cosmic medium is now regarded as a matter which particularly
is in the BEC state and its particle number current and
energy-momentum tensor have the form (\ref{sfPN}) and
(\ref{sfEMT}). We assume that the self-dependent condensate ansatz
(\ref{AnsatzSeparatedCondensate}) holds and the superfluid
background obeys the equation of state (\ref{ChaplyginEOS}) and
the excited state is described by the relations (\ref{EOSnn}) and
(\ref{AnsatzPn}). This means that we ignore nonlinear high-energy
part of the quasiparticle spectrum. Although these restrictions
render the model incomplete they essentially simplify the
evolution equations retained simultaneously the key features of
the model.

Let us consider a homogeneous and isotropic spatially flat
universe. In this case the superfluid and normal velocities are
equal and thus $\gamma=1$. Einstein equations then reduce to
\beq\label{EinsteinEqs}
3\frac{\dot a^2}{a^2} = 8\pi G \rho_{\textrm{tot}},
\qquad
-6\frac{\ddot a}{a} = 8\pi
G(3p_{\textrm{tot}}+\rho_{\textrm{tot}}),
\eeq
where $\rho_{\textrm{tot}}$ consists of the condensate density
$\rho_{\textrm{c}}$ and the normal one
$\rho_{\textrm{n}}=W_{\textrm{n}}-p_{\textrm{n}}$ that are
interpretable as DE and DM densities respectively, and
$p_{\textrm{tot}}=p_{\textrm{c}}+p_{\textrm{n}}$. In accordance
with the integrability conditions of Einstein equations we require
local energy-momentum conservation $\nabla_\mu T^{\mu\nu}=0$ that
yields
\beq\label{EnergyConservation}
\dot\rho_{\textrm{tot}} + 3\frac{\dot
a}{a}(p_{\textrm{tot}}+\rho_\textrm{tot})=0.
\eeq
The interaction between DE and DM is implicitly included in
equation (\ref{EnergyConservation}) and also in particle number
conservation $\nabla_\mu n^\mu = 0$ that leads to
\beq\label{ParticleConservation}
\dot n_{\textrm{tot}} + 3\frac{\dot a}{a}n_{\textrm{tot}}=0
\qquad\Longrightarrow\qquad%
n_\textrm{c}+n_\textrm{n} = \frac{n_0}{a^3}, \quad
n_0=\mbox{const}.
\eeq
This approach distinguishes the present model from
\cite{Morikawa1} where the rate of the transition between ground
and excited states $\Gamma$ is explicitly used as an interaction
factor and equation (\ref{EnergyConservation}) breaks down into
separated balance equations for DE and DM. In \cite{Zhang1} the
similar splitting is applied  for interaction between Chaplygin
gas (it is regarded as DE) and CDM.

Taking into account the expressions
(\ref{EOSnn})--(\ref{AnsatzPn}) and (\ref{ParticleConservation})
we reduce equations (\ref{EinsteinEqs}) and
(\ref{EnergyConservation}) to following two dimensionless
equations:
\beqar
&&%
 3(1+\polytrop)\frac{\dot a^2}{a^2} = \frac{1}{\rho}+
\frac{k}{a^3}\left(\frac{\polytrop\rho}{\sqrt{\rho^2-1}}+\frac{\sqrt{\rho^2-1}}{\rho}\right),
\label{Eq1}\\&&%
3\frac{\dot a}{a}\left(1+\polytrop-
            \frac{k}{a^3}\frac{1}{\sqrt{\rho^2-1}}\right) +
            \frac{\dot\rho}{\rho}
            \left(1- \frac{k}{a^3}\left(\frac{1}{\sqrt{\rho^2-1}}-
            \frac{\polytrop\rho^2}{(\rho^2-1)^{3/2}}\right) \right)=0,
\label{Eq2}
\eeqar
where $\rho=\rho_{\rm c}/2M$, and $k=n_0/2\sqrt{\lambda M}$. The
dimensionless time variable $t'$ is connected with real time $t$
as $t'=\sqrt{16\pi G M} t$.

\subsection{Exact solution}

In the formal limit $\polytrop\to\infty$ equations (\ref{Eq1}) and
(\ref{Eq2}) are solved analytically. As obvious from
(\ref{AnsatzPn}) the quasiparticle pressure is neglected and DM
behaves as dust-like matter. In this case Eq.~(\ref{Eq2}) yields
\beq
\rho_{\rm c} = \sqrt{\frac{k^2}{(a^3+b^3)^2}+1},
\eeq
and the scale factor varies in according to the integral
\beq
t'=\int \frac{\sqrt{3a}\d
a}{\sqrt{a^3+b^3}\sqrt[4]{k^2(a^3+b^3)^{-2}+1}}.
\eeq

The parameter $k$ gives an initial normalized total particle
number density and $b$ associates with an initial particle number
density for the normal component. More precisely, $b^3
=(n_0/n_{\textrm{n}}(0)-1)^{-1}$. If $b=0$, the normal component
is unavailable and the evolution follows the scenario proposed in
\cite{Kamenshchik1} since the ground state obeys the equation of
state (\ref{ChaplyginEOS}). In the case of $b\ne 0$ DM is governed
by the law
\beq
\rho_{\textrm{n}} = \frac{b^3}{a^3}
\sqrt{\frac{k^2}{(a^3+b^3)^2}+1}.
\eeq

At the beginning stage (i.e. for small $a$) the total energy
density is approximated by $\rho_{\textrm{tot}}\propto a^{-3}$
that corresponds to a universe dominated by dust-like matter. The
same behavior is a feature of  Chaplygin gas \cite{Kamenshchik1}
but even though in this model the condensate has the same equation
of state, such  dependence is due to the normal component.

At the late stage (i.e. for large $a$) $\rho_{\textrm{tot}}\to 1$.
Separating now DE and DM contributions one finds the subleading
terms are
\beqar
\rho_{\rm c} & \sim & 1+\frac{k^2}{2}a^{-6}, \label{asymptotChG} \\
\rho_{\textrm{n}} & \sim & \frac{b^3}{a^3},\label{asymptotDust}
\eeqar
whereas the scale factor time evolution corresponds to de Sitter
spacetime, namely, $a\propto e^{t'/\sqrt{3}}$. At this stage the
fluid is almost in the ground state so that the condensate wave
function is close to the potential minimum. The behavior is
similar to GCG \cite{Bento1}: expressions (\ref{asymptotChG}) and
(\ref{asymptotDust}) imply that the system evolves as a mixture of
the cosmological constant and the dust-like matter. Note, the
asymptotic formula (\ref{asymptotChG}) is valid for any value of
$\polytrop$.

\subsection{Numerical simulation}

When $\polytrop$ has a finite value equations (\ref{Eq1}) and
(\ref{Eq2}) are solved numerically. It emerges that in the context
of the concerned scenario the universe expansion may be
decelerated or accelerated from the start. Nevertheless once the
universe starts accelerating it cannot decelerate anymore and
eventually falls within de~Sitter phase. We regard only the
solutions with initial deceleration. Photometric observations of
apparent Type Ia supernovae attests that the recent cosmological
acceleration commenced at $0.3<z<0.9$ \cite{Wang1}. To fix time
scaling, $\dot a$ is assumed to be in the minimum when the
redshift $z=0.5$.

\begin{figure}[!tbh]
\includegraphics[width=.45\textwidth]{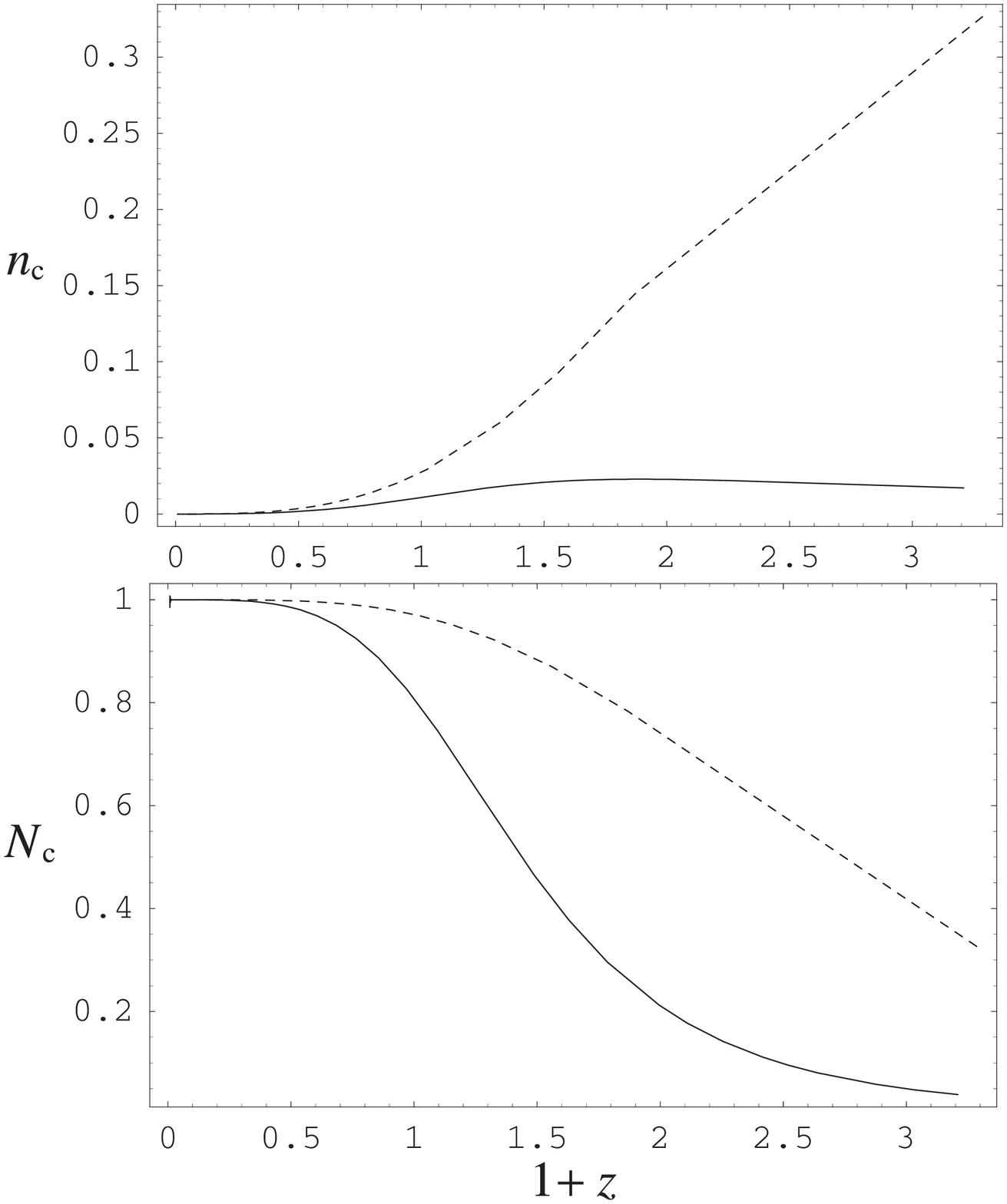}%
\hspace{.1\textwidth}
\parbox[b]{.45\textwidth}{\caption{The condensate number density in the units of $n_0$ as
a function of the redshift $z$ for $k=15$ and $\polytrop=3$. The
solid ($\rho(0)=1.02$) and dashed ($\rho(0)=5$) curves correspond
to deficit and exceed of the particles in the ground state
respectively}\label{FigN}}\\
\vspace{2\baselineskip}\\
\includegraphics[width=.45\textwidth]{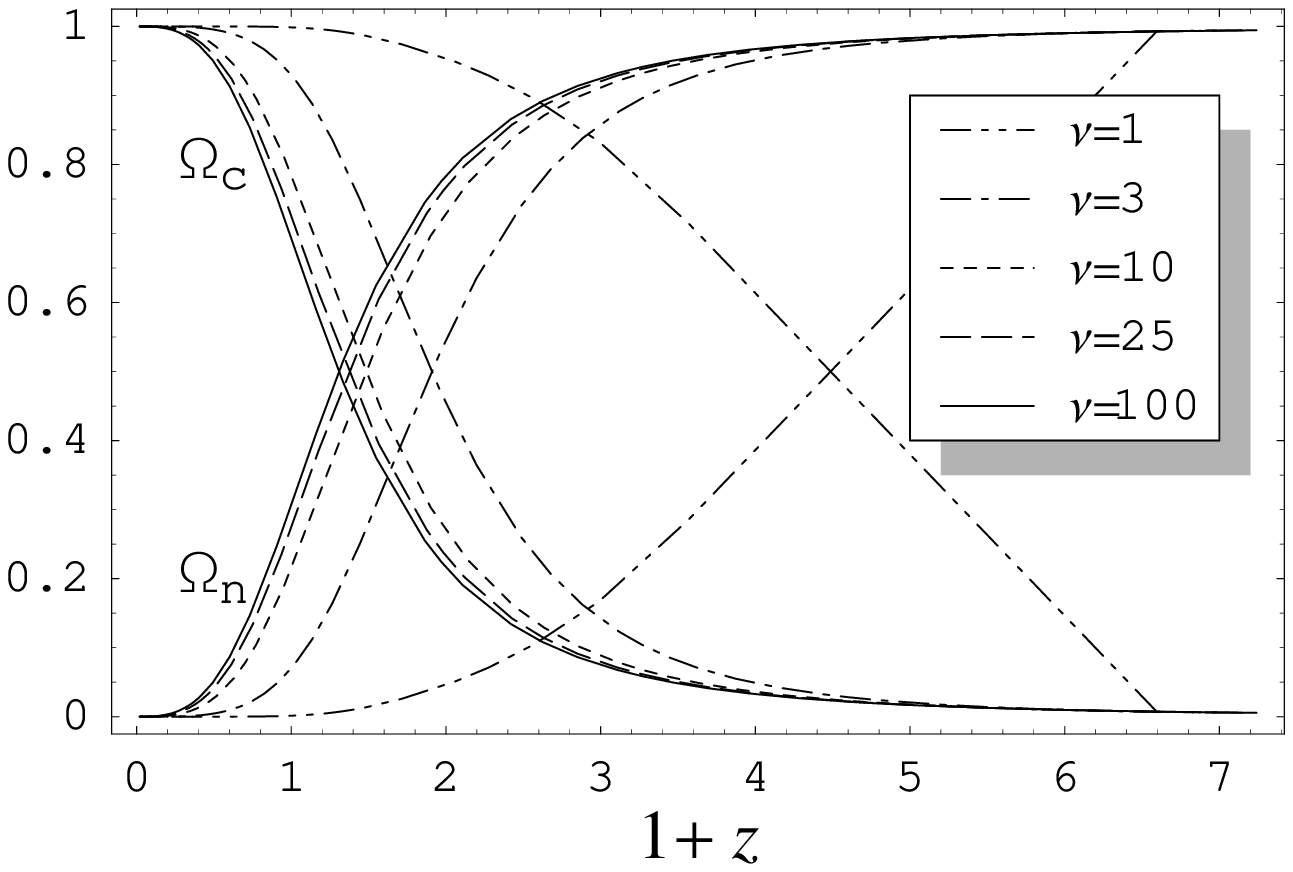}%
\hspace{.1\textwidth}
\parbox[b]{.45\textwidth}{\caption{The ratio of the energy density to the critical density
for the different components as a function of the redshift $z$ for
$k=120$ and $\rho(0)=1.2$}\label{FigOmega}}\\
\vspace{2\baselineskip}\\
\includegraphics[width=.45\textwidth]{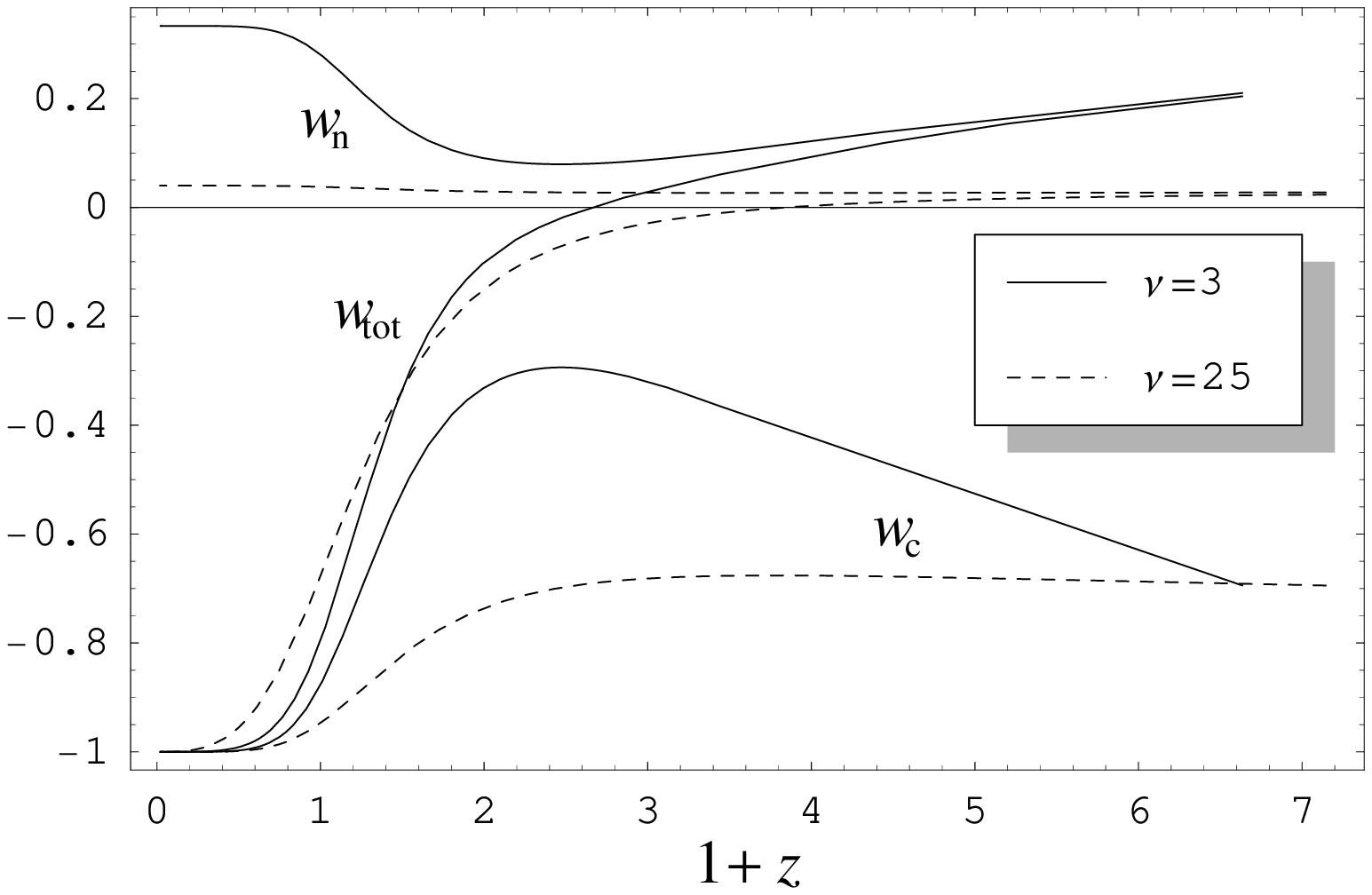}%
\hspace{.1\textwidth}
\parbox[b]{.45\textwidth}{\caption{The equation-of-state parameters as a function of the
redshift $z$ for $k=120$ and $\rho(0)=1.2$}\label{FigW}}
\end{figure}

Equations (\ref{Eq1}) and (\ref{Eq2}) are solved with initial
conditions $a(0)=1$ and $\rho(0)$ is a constant greater than 1.
The coefficient $k$ is imposed to be more than
$\sqrt{\rho(0)^2-1}$ to ensure that both DE and DM contents are
available . At the final stage DM dies out gradually. This means
that all particles pass into the ground state and the normalized
condensate number of particles in a comoving volume $a^3$, namely
$N_{\rm c}=n_{\rm c}a^3$, goes to 1 while the number density
scalar $n_{\rm c}$ decays to zero (see Fig.~\ref{FigN}). If we
start with a small value of the background number density then the
beginning condensate production rate is so high that not only the
condensate particle number $N_{\rm c}$ increases but the number
density $n_{\rm c}$ as well. To the contrary, an excess of the
condensate particles involves monotone behavior of the condensate
number density and the particle number in the comoving volume.

Fig.~\ref{FigOmega} depicts an evolution of the normalized energy
densities $\Omega_{\rm c}$ and $\Omega_{\rm n}$ of DE and DM
respectively. The curves are plotted for different values of
$\polytrop$ and demonstrate increasing of the DM content when
$\polytrop$ increases. Correspondence with the current
observational value of the DE fraction $\Omega_{\rm c}\approx
0.72$ falls on $\polytrop\approx 25$. This implies a high-degree
temperature dependence of the phonon energy and the second sound
speed with a value much less than it would be expected for real
phonons. Note, that in superfluid helium a lower second sound
speed is provided by quasiparticles from the nonlinear part of the
energy spectrum (such as rotons). In the context of the pure
phonon consideration they are not taken into account and their
influence is simulated with a large value of $\polytrop$.

Plots of equation-of-state parameters $w_{\rm A}=p_{\rm
A}/\rho_{\rm A}$ (A=c, n, tot)  as a function of the redshift $z$
is shown  on Fig.~\ref{FigW}. The DE equation-of-state $w_{\rm c}$
occurs close to $-1$ extremely fast after the universe starts to
accelerate. This is accompanied by vanishing of the excitations
and the total equation-of-state $w_{\rm tot}$ is found to be the
same at the same time. Simultaneously the DM one $w_{\rm n}$
approaches the value of $1/\polytrop$ that is the asymptotic value
for the second sound speed as the sound speed $c_{\rm s}\to 1$. If
$\polytrop$ is large then DM is perceived as a nonrelativistic
one. This CDM-like behavior is observed within the whole range of
the redshift under consideration. It is expected that the
excitations with a nonlinear dispersion will promote decreasing
the equation-of-state $w_{\rm n}$ when the complete quasiparticle
spectrum is taken into account.

\section{Conclusion}

In this letter we examine the model of superfluid Chaplygin gas
(SCG) describing the dark sector of the universe as  a matter that
behaves as DE while it is in the ground state and as DM when it is
in the excited state. Cosmological dynamics is described in the
framework of the
two-fluid model therefore %
the interaction between DE and DM is implicitly involved into the
conservation laws (\ref{EnergyConservation}) and
(\ref{ParticleConservation}). In this approach there is no need to
introduce different equations for the description of DE and DM
evolution and to use the interaction factor as an additional
parameter. Moreover, if we abandon the self-dependent condensate
ansatz (\ref{AnsatzSeparatedCondensate}), it will be impossible to
uniquely divide the total energy density into the DE and DM
fractions.

The SCG model is applied to the universe evolution from the
deceleration-acceleration transition epoch. It provides the
current mixture of the dark contents and approaches de Sitter
phase in the future. The normal component (DM) is formed by a pure
phonon gas  with the pressure varying as the power $\polytrop+1$
of temperature. Simple fitting the model parameters to the
observational data shows that $\polytrop$ must be quite large. In
this case DM behaves as CDM. In the standard model CDM consists of
nonrelativistic massive particles whereas in this model it is
described as an excited state of an exotic matter with a low
second sound speed. %
To develop more realistic model, a wide quasiparticle spectrum
should be taken into account %
so that the quasiparticles with a nonlinear dispersion are
dominant in the modern epoch while the phonons prevail in the
future at the temperatures close to zero. Further all the
particles fall into the ground state and the final epoch is de
Sitter universe. The model is integrated in more early epochs as a
usual matter since it almost fully is in the excited state that
time.

It should be emphasized the difference between the GCG and SCG
models. In the former the energy density $\rho_{\rm tot}=\rho_{\rm
Ch}$ consists of both vacuum and matter contributions. The
parameter $\alpha$ (see introduction) is assumed to satisfy the
constraint $\alpha<0.4$ \cite{Barreiro1} to be in agreement with
observations. In the latter $\rho_{\rm tot}=\rho_{\rm c}+\rho_{\rm
n}$, where the condensate with the density $\rho_{\rm c}=\rho_{\rm
Ch}$ can be considered as GCG and the previous restriction is
unnecessary. Moreover ordinary Chaplygin gas ($\alpha=1$) is
preferred since $\alpha$ has no such a dramatic effect for this
model as in \cite{Bento1,Bilic1,Bento2} and its governing role for
the interaction between DE and DM goes to $\nu$. This is not to
say that $\alpha$ and $\nu$ have an identical significance. We can
see that while the GCG model becomes equivalent to $\Lambda$CDM
for $\alpha=0$, SCG persists as a model unified DE and DM for
$\nu\to\infty$. Therefore we can use $\alpha$ as an additional
parameter in the SCG model later.

There is another problem in the models of DE and DM unification.
It concerns the non-negligible sound speed that produces
unphysical oscillations and an exponential blow-up in the DM power
spectrum at present \cite{Sandvik1}. This problem was solved for
the GCG model in \cite{Bento2} by the special decomposition of the
energy density into DE and DM components. In the SCG model DM is
generated by excitations and a role of the sound is played by the
second sound. We should expect that a perturbative analysis of the
energy density fluctuations of DM will lead to additional
restrictions on the parameters. The heuristic estimation of
$\alpha$ relying on scales of Galaxy clusters obtained in
\cite{Sandvik1} gives the range $|\alpha|<10^{-5}$ for GCG that
means its indistinguishability from $\Lambda$CDM. The similar
estimation for SCG can be given by the formal replacement
$\alpha\to 1/(1+\nu)$. As one would expect pressureless DM
($\nu\to\infty$) is free from the blow-up and the restriction
$\nu>10^{5}$ is in agreement with the foregoing inference
regarding the quasiparticle spectrum.

\newpage

\end{document}